**Sign change in *c*-axis thermal expansion and lattice collapse by Ni substitution in $Co_{1-x}Ni_xZr_2$ superconductors**

Yuto Watanabe[1], Hiroto Arima[1], Hidetomo Usui[2], Yoshikazu Mizuguchi[1]*

1. Department of Physics, Tokyo Metropolitan University, 1-1, Minami-osawa, Hachioji 192-0397, Japan.
2. Department of Physics and Materials Science, Shimane University, Matsue, Shimane 690-8504, Japan

**Abstract**

We investigated the structural, electronic, and superconducting properties of $Co_{1-x}Ni_xZr_2$, in which *c*-axis thermal expansion is systematically controlled. At $x \leq 0.3$, *c*-axis negative thermal expansion (NTE) was observed, and the thermal expansion constant $\alpha_c$ approached zero with increasing $x$. At $x = 0.4$–$0.6$, zero thermal expansion was observed, and positive thermal expansion (PTE) appeared for $x \geq 0.7$. By analyzing the $c/a$ ratio, we observed a possible collapsed transition in the tetragonal lattice at around $x = 0.6$–$0.8$. The lattice collapse results in *c*-axis PTE and the suppression of bulk superconductivity.

**Keywords**: negative thermal expansion, superconductor, $Co_{1-x}Ni_xZr_2$, lattice collapse transition

**Impact Statement**

Systematic Ni substitution in the $Co_{1-x}Ni_xZr_2$ superconductor achieves sign change in *c*-axis thermal expansion constant and collapsed tetragonal phases; those structural changes are linked to suppression of superconductivity.

**1. Introduction**

Thermal expansions are structural properties of materials. In the case of normal (positive) thermal expansion (PTE), an axis and/or volume expand with increasing temperature. In contrast, materials with negative thermal expansion (NTE), those contract with increasing temperature. Importantly, zero thermal expansion (ZTE) can be achieved by fabricating a composite using PTE and NTE materials, and the ZTE materials have been used in various structural materials and devices in which ultraprecision of positions is required [1−5]. However, achievement of ZTE in a single material is quite rare [6] but has potential merits for development of ZTE application. Development of ZTE in a superconductor is particularly interesting because it will be available in superconducting devices like Josephson junctions with a strength to temperature cycle.

Recently, we reported anomalous axis thermal expansion in $CuAl_2$-type (tetragonal) transition-metal zirconide superconductors $TrZr_2$ ($Tr$: transition metal) [7–9]. In $CoZr_2$, for example, the $c$-axis shows NTE in a wide temperature range, while the $a$-axis exhibits PTE. Owing to the contrasting axis thermal expansion, $CoZr_2$ and similar $TrZr_2$ show volume ZTE in a limited temperature range. In addition, we revealed that the axis ratio $c/a$ is the potential factor for switching the character of the $c$-axis expansion [9]. In this study, we focus on $CoZr_2$ and $NiZr_2$ with a large and small $c/a$ ratio, respectively. $CoZr_2$ exhibits a $c$-axis NTE and is a superconductor with a transition temperature ($T_c$) of ~6 K [7,10,11]. $NiZr_2$ exhibits PTE in both $a$ and $c$ axes. In previous works [12−14], synthesis and physical properties of a solid solution system of $Co_{1-x}Ni_xZr_2$ were reported with its superconducting properties. Here, we show that the $c$-axis thermal expansion character in $Co_{1-x}Ni_xZr_2$ is systematically changed from NTE, ZTE, and PTE with increasing Ni concentration $x$.

**2. Methods**

Polycrystalline samples of $Co_{1-x}Ni_xZr_2$ ($x$ = 0, 0.1, 0.2, 0.3, 0.4, 0.5, 0.6, 0.7, 0.8, 0.9, 1.0) were synthesized by arc melting in an Ar atmosphere. Powders of pure transition metals ($Tr$) of Co (99%, Kojundo Kagaku) and Ni (99.9%, Kojundo Kagaku) with a nominal composition were mixed and pelletized. The $Tr$ pellet and plates of pure Zr (99.2%, Nilaco) were used as starting materials. The samples were melted five times and turned over after melting to homogenize the sample.

X-ray diffraction (XRD) patterns were collected by $\theta$-$2\theta$ method with Cu-K$\alpha$ radiation on a Miniflex-600 (RIGAKU) diffractometer equipped with a high-resolution semiconductor detector D/tex-Ultra. For High-temperature XRD on a Miniflex-600, the sample temperature was controlled by a BTS 500 attachment. The obtained XRD patterns were refined by the Rietveld method using RIETAN-FP [15], and the schematic images of the crystal structure were depicted using VESTA [16]. The actual compositions of the samples were investigated using energy-dispersive x-ray spectrometry (EDX, Swift-ED, Oxford) on a scanning electron microscope (SEM, TM3030, Hitachi Hightech).

The temperature dependence of magnetization was measured both after zero-field cooling (ZFC) and field cooling (FC) using a superconducting quantum interference device (SQUID) on an MPMS3 (Quantum Design).

The first principles band calculations were performed using the WIEN2k package [17] within the PBE-GGA exchange-correlation functional [18]. The virtual crystal approximation is adopted to take into account the effect of the elemental substitution of Ni for Co. We used the experimentally determined lattice parameters shown in Table 1. The atomic coordinates of Zr were theoretically optimized. $RK_{max}$ and the $k$-mesh were set to 8 and 10×10×10, respectively.

## 3. Results and Discussion
### 3-1. Crystal structure analysis and axis thermal expansion

The obtained actual compositions at the $Tr$ site are comparable to the nominal values and summarized in Table 1. Figure 1(a) shows the schematic images of crystal structure of $Co_{1-x}Ni_xZr_2$. Figures S1(a)-S1(c) (supporting materials) are XRD patterns for $x$ = 0–1.0. These compounds have a tetragonal CuAl$_2$-type structure ($I4/mcm$), and the main peaks could be indexed with the structural model. Small impurity peaks of the orthorhombic $Tr$Zr$_3$ phase are seen as indicated by asterisks as reported in Ref. 7, and the amount of the impurity decreased with increasing Ni concentration (*x*). We estimated lattice constants by Rietveld refinements, and the obtained parameters are plotted in Fig. 1(b) and summarized in Table 1. The obtained trend of lattice constants is consistent with a previous study [14].

Table 1. Results of chemical analyses, evolutions of lattice constants and thermal expansion constants, and $T_c$ in examined $Co_{1-x}Ni_xZr_2$. $T_c$ with bracket indicates filamentary superconductivity.

| Nominal *x* | *x* (EDX) | *a* (Å) at 303 K | *c* (Å) at 303 K | *c/a* at 303 K | $\alpha_a$ (μK$^{-1}$) | $\alpha_c$ (μK$^{-1}$) | $\beta$ (μK$^{-1}$) | $T_c$ (K) |
|---|---|---|---|---|---|---|---|---|
| 0 | 0 | 6.360(3) | 5.514(3) | 0.8670(6) | 25.6(6) | -20(1) | 31(1) | 5.9 |
| 0.1 | 0.111(3) | 6.375(3) | 5.475(3) | 0.8589(6) | 20.2(8) | -14(1) | 27(2) | 6.4 |
| 0.2 | 0.216(6) | 6.3750(9) | 5.428(1) | 0.8514(2) | 21.1(6) | -4.6(9) | 37(1) | 6.1 |
| 0.3 | 0.312(2) | 6.377(3) | 5.391(2) | 0.8454(5) | 14.9(6) | -3(1) | 27(1) | 5.1 |
| 0.4 | 0.417(3) | 6.415(1) | 5.384(2) | 0.8393(3) | 14.0(7) | -1(1) | 27(2) | 4.1 |
| 0.5 | 0.544(6) | 6.425(1) | 5.350(1) | 0.8327(3) | 15.1(5) | 1.4(9) | 32(1) | 3.2 |
| 0.6 | 0.635(5) | 6.450(2) | 5.336(3) | 0.8273(8) | 23.6(7) | 2(2) | 48(2) | 2.5 |
| 0.7 | 0.752(4) | 6.469(3) | 5.288(3) | 0.8175(6) | 14.4(9) | 15(2) | 43(3) | (2.4) |
| 0.8 | 0.824(2) | 6.481(3) | 5.279(3) | 0.8145(6) | 7(1) | 11(2) | 25(3) | (2.5) |
| 0.9 | 0.937(3) | 6.491(4) | 5.261(3) | 0.8105(7) | 10.5(6) | 12(1) | 33(2) | (2.5) |
| 1 | 1 | 6.509(6) | 5.259(4) | 0.8081(9) | 15.8(5) | 17(1) | 49(1) | (2) |

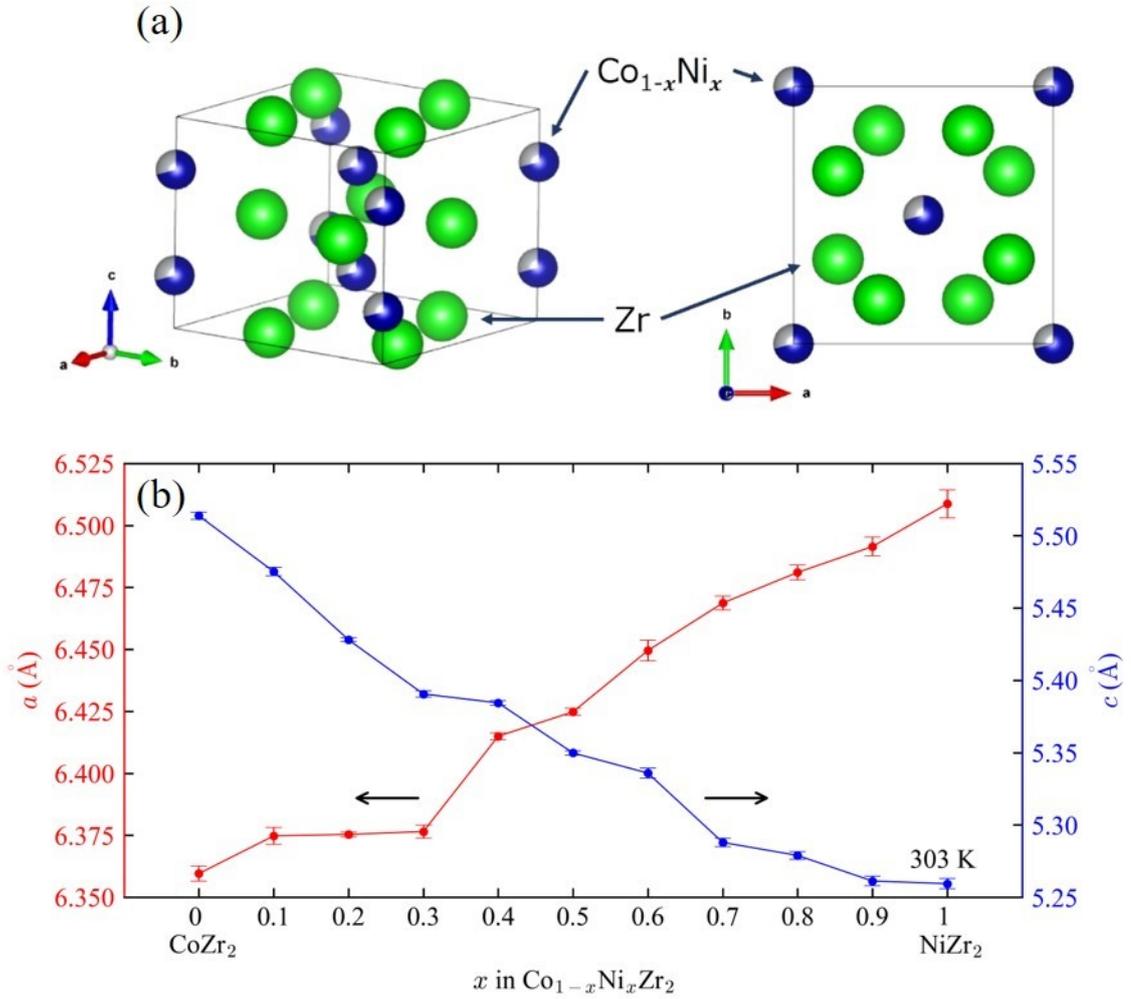

Figure 1. (a) Schematic images of the crystal structure of $Co_{1-x}Ni_xZr_2$. (b) Ni concentration ($x$) dependence of lattice constants $a$ and $c$ at 303 K. The error bars are standard deviations estimated by the Rietveld refinement.

We investigated the thermal expansion properties by high-temperature XRD. As reported in Ref. 7, $CoZr_2$ exhibits $c$-axis NTE, while the $a$-axis exhibits PTE. To investigate how the anisotropic axis thermal expansion changes with Ni doping, we collected the data for all the samples between 303-573 K with a temperature increment of about 30 K and used it for the calculation of the linear thermal expansion coefficients along the $a$-axis ($\alpha_a$) and $c$-axis ($\alpha_c$) and the volumetric expansion coefficient ($\beta$). The $c$-axis NTE was observed for $x$ = 0, 0.1, 0.2, 0.3; hence, we show the results for $x$ = 0.3 as an example. Figures S2(a)-S2(c) (supporting materials) show the temperature dependence of lattice constants $a$, $c$, and volume ($V$) for the $x$ = 0.3 sample.

Figure S2(d) (supporting materials) shows the typical high-temperature XRD patterns. The 002 peak shifts to the higher angle side with increasing temperature, indicating that the $x = 0.3$ sample still contracts along the $c$-axis upon heating. The estimated values of $\alpha_a$, $\alpha_c$, and $\beta$ using the formulas $\alpha_a = \frac{1}{a\,(303\,\text{K})} \cdot \frac{da}{dT}$, $\alpha_c = \frac{1}{c\,(303\,\text{K})} \cdot \frac{dc}{dT}$, and $\beta = \frac{1}{V\,(303\,\text{K})} \cdot \frac{dV}{dT}$ are $\alpha_a = +14.9(6)$, $\alpha_c = -3(1)$, $\beta = +27(1)$ µK$^{-1}$, respectively The magnitude of $\alpha_c$ for $x = 0.3$ is smaller than that of CoZr$_2$, which also exhibits NTE along the c-axis with $\alpha_c < -15$ µK$^{-1}$ [7]. This suggests that the substitution of Ni for the Co site suppresses the NTE along the $c$-axis, and the switching between PTE and NTE is controlled by adjusting the $x$ value. Figures 2(a)-(i) show the temperature dependence of the normalized rate of change in the lattice constants $a$, $c$, and $V$ from 303 K for all the samples. For all $x$, the lattice constant $a$ and $V$ gradually increase with heating. The samples with a lower Ni amount ($x = 0$–0.3) show NTE along the $c$-axis as shown in Fig. 2(d). On the other hand, the samples with a larger Ni amount ($x = 0.7$–1) show PTE along the $c$-axis as shown in Fig. 2(f). For the samples with medium Ni amount ($x = 0.4$–0.6), ZTE trend was observed as shown in Fig. 2(e). Figure 3 shows the $x$ dependence of the linear thermal expansion coefficient along the $c$-axis, which shows a successful control of the switching of NTE and PTE along the $c$-axis by tuning $x$. The turning point for the NTE and PTE is estimated between $x = 0.4$ and 0.6. Therefore, there is a possibility to synthesize the sample which exhibits the perfect ZTE along the $c$-axis by optimizing the Ni amount doped at the Co site. As well, materials that exhibit anisotropic thermal expansion have been reported, such as β-Eucryptite (LiAlSiO$_4$) [19], Ag$_3$[Co(CN)$_6$] [20], and Ca$_2$RuO$_4$ [21]. The mechanisms of NTE are diverse [1,22]. For example, the cause of the NTE on the monoclinic Ca$_2$RuO$_4$ is d$_{xy}$ orbital ordering and disordering [21]. Not only electronic contributions but also structural properties contribute to the NTE mechanisms. In another study on α-(Cu$_{2-x}$Zn$_x$)V$_2$O$_7$, the chemical substitution of Cu by Zr decreases the free space for the transverse vibrations, which suppresses NTE along the $b$-axis [23]. Furthermore, Mn$_3$Cu$_{1-x}$Ge$_x$N exhibits giant negative thermal expansion due to the local lattice distortion triggered by Ge dope [24]. These facts will help us to understand the mechanisms of the anomalous (anisotropic) $c$-axis thermal expansion in the current system. Recently, we reported that the NTE along the $c$-axis for $Tr$Zr$_2$ was caused by the robust $Tr$-Zr distance to the temperature change and the flexible bonding of the $Tr$Zr$_8$ polyhedron units. In addition, the $c/a$ ratio is found to be an essential parameter that determines the polyhedron shape and the thermal expansion characteristics [7,9]. Therefore, further studies on electronic and/or orbital characteristics and local structures of Co$_{1-x}$Ni$_x$Zr$_2$ will be striking in determination of the mechanisms of the emergence of $c$-axis NTE in $Tr$Zr$_2$.

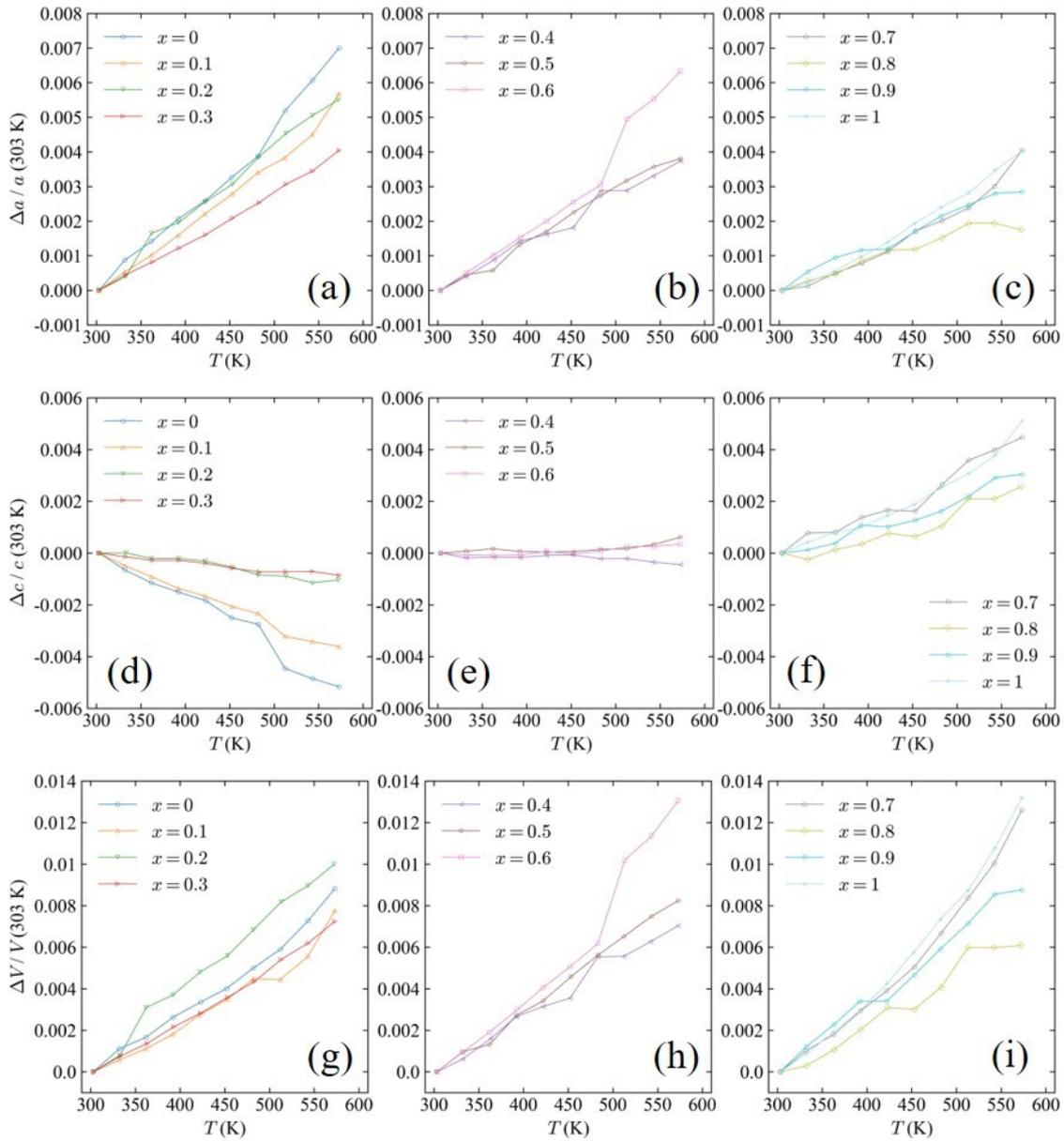

Figure 2. (a-i) Temperature dependence of the normalized rate of change of the lattice constants $a$, $c$, and $V$ from 303 K.

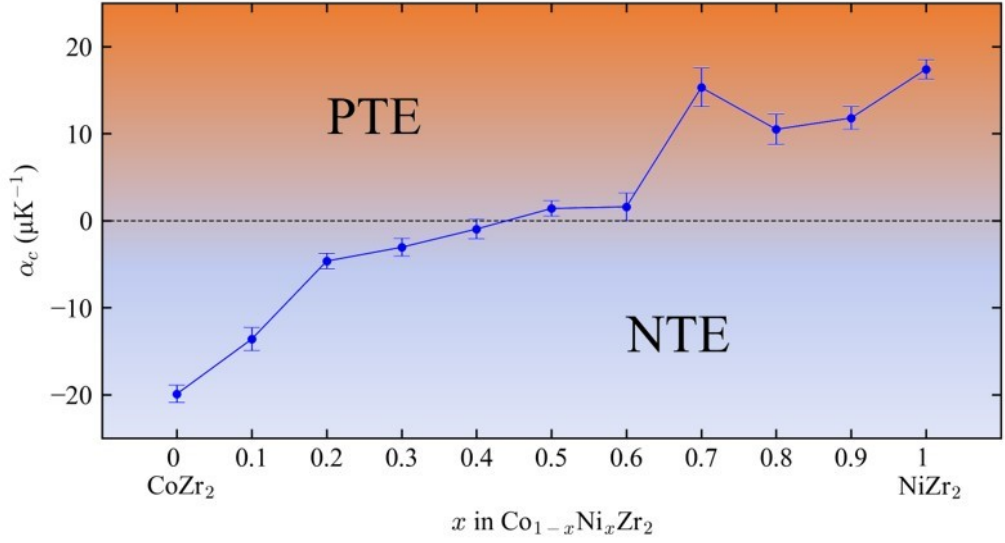

Figure 3. The $x$ dependence of $\alpha_c$ for $Co_{1-x}Ni_xZr_2$.

### 3-2. Superconducting properties

A superconducting transition was observed for all $Co_{1-x}Ni_xZr_2$ samples, as shown in Fig. 4(a). The large diamagnetic signals observed for $x \leq 0.6$ suggests the emergence of bulk superconductivity. In contrast, the signals for $x > 0.6$ are quite small, which indicates that the observed diamagnetic signals are caused by filamentary (trace) superconductivity states in those samples. The $T_c$ tends to decrease with increasing $x$; however, the samples with $x = 0.1$ and 0.2 have $T_c$ slightly higher than that for $x = 0$. Figure 4(b) shows the enlarged view near the $T_c$ for = 0, 0.1, 0.2, 0.3, 0.4. The highest $T_c$ of 6.39 K was observed for $x = 0.1$. To discuss about the electronic origins on this behavior, we performed first-principles calculations for $Co_{1-x}Ni_xZr_2$. Figure 5(a) shows the $x$ dependence of density of states near Fermi level, DOS($E_F$). The $x$ dependence of the calculated DOS($E_F$) looks consistent with the evolution of $T_c$ if we assumed conventional phonon-mediated superconductivity [25], because a large DOS($E_F$) achieves a higher $T_c$ in a conventional superconductor. Figure 5(b) shows the $x$ dependence of $T_c$. As we mentioned above, the evolution of $T_c$ at $x = 0.1$ is consistent to the DOS($E_F$) behavior where the $x$ value is smaller than $x = 0.7$. However, we cannot explain the change in $T_c$ with DOS($E_F$) behavior where $x$ is larger than $x = 0.7$. As mentioned above, the samples with $x \geq 0.7$ exhibit filamentary superconductivity; in Fig. 5(b), we indicated the boundary between bulk superconductivity (Bulk

SC) and filamentary superconductivity (Filamentary SC). According the discussion above, we suggest that the bulk superconductivity observed for $x \leq 0.6$ is positively linked to DOS($E_F$), which would suggest the importance of phonon in the superconductivity mechanism in $Tr$Zr$_2$.

For $x \geq 0.7$, bulk superconductivity is suppressed, while the DOS($E_F$) is comparable or higher than $x = 0.6$. To explore possible cause of the suppression of superconductivity, we estimated the $c/a$ ratio of Co$_{1-x}$Ni$_x$Zr$_2$ using the data at 303 K and plotted in Fig. 5(c) as a function of $x$. Although $c/a$ linearly decreases with increasing $x$ for $x \leq 0.7$, the slope clearly changes at around $x = 0.6-0.8$. For $x = 0.7-1.0$, another slope can guide the evolution of $c/a$. We propose that the change in the $c/a$ ratio is a kind of transition to collapsed tetragonal phases as observed in iron-based superconductors CaFe$_2$As$_2$ and KFe$_2$As$_2$ and related layered compound [26-30]. The electronic structure is generally affected by a collapsed tetragonal transition, which affects superconductivity as well [31,32], we assume that the disappearance of bulk superconductivity by Ni heavy doping is related to the collapsed transition. In our previous work, we suggested the trend that a higher $T_c$ of $Tr$Zr$_2$ is achieved with increasing lattice constant $c$ [33]. This fact is also consistent with the above scenario because the $c$-axis is largely compressed at around $x = 0.7$. To obtain further evidence on the collapsed transition and its relation to electronic structure, superconductivity, and axis thermal expansion, further investigations with different probes are needed.

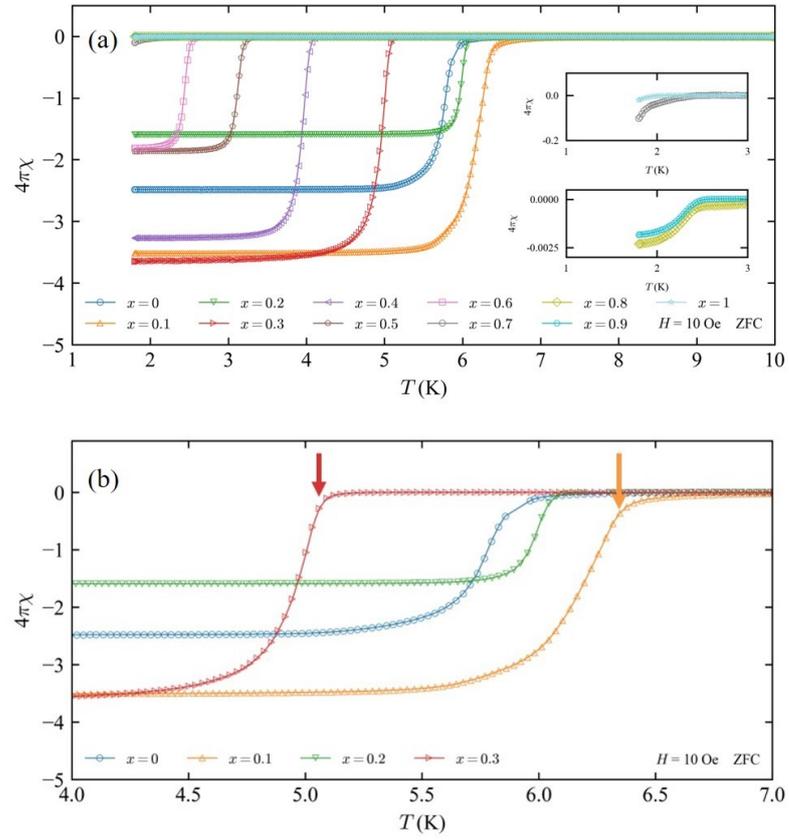

Figure 4. (a) Temperature dependence of ZFC susceptibility for $Co_{1-x}Ni_xZr_2$. (b) Enlarged view near the superconducting transition temperature for $x$ = 0, 0.1, 0.2, 0.3, 0.4.

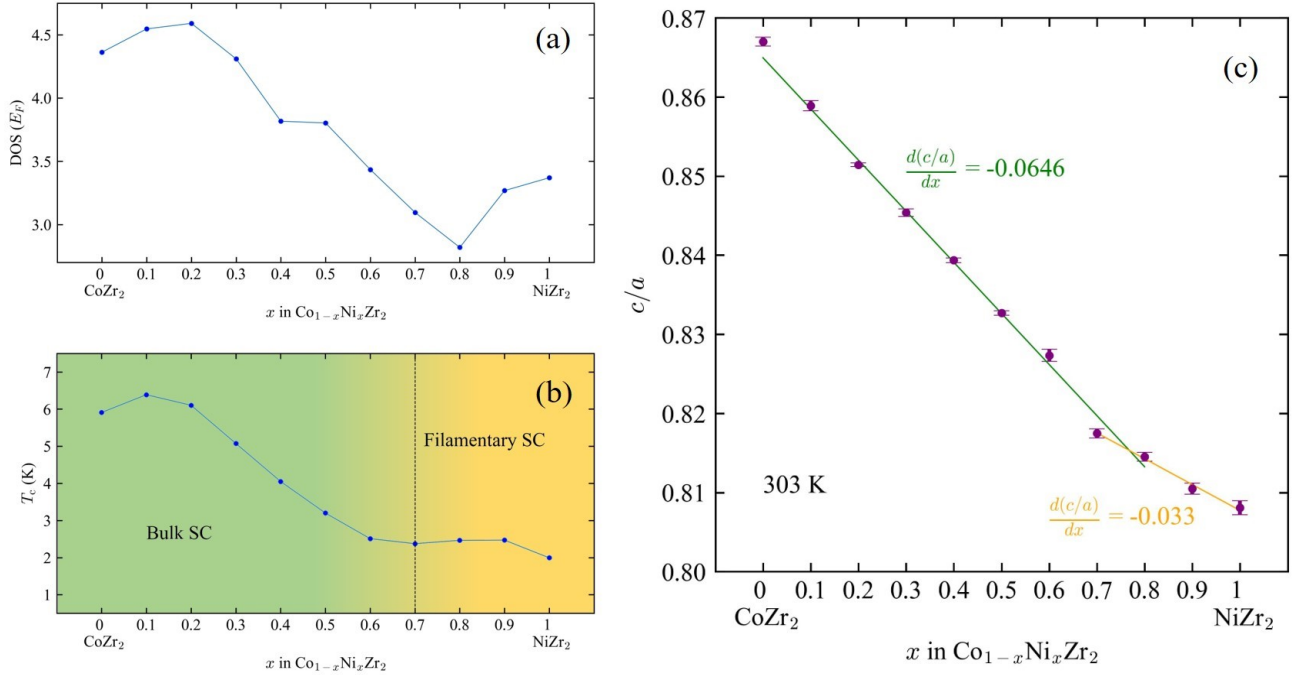

Figure 5. (a) The $x$ dependence of DOS($E_F$) for $Co_{1-x}Ni_xZr_2$. (b) The $x$ dependence of superconducting transition temperature for $Co_{1-x}Ni_xZr_2$. (c) The $x$ dependence of $c/a$ for $Co_{1-x}Ni_xZr_2$ at 303 K. The estimated slope values for $x = 0 – 0.6$ and $0.7 – 1$ are $-0.0646 \pm 0.0009$, $-0.033 \pm 0.003$ respectively.

## 4. Conclusion

We investigated the crystal structure, axis thermal expansion, electronic structure, and superconducting properties of transition-metal zirconide superconductor $Co_{1-x}Ni_xZr_2$. The samples were synthesized by arc melting and characterized by powder XRD and EDX. At $x \leq 0.3$, $c$-axis NTE was observed, and the thermal expansion constant ($\alpha_c$) approached zero with increasing $x$. At $x = 0.4–0.6$, $c$-axis thermal expansion close to ZTE was observed, and PTE appeared for $x \geq 0.7$. Those results confirm that the $c$-axis NTE can be controlled by Ni substitution (tuning $c/a$ ratio) and switched to PTE. On the superconducting properties, we observed bulk superconductivity for $x \leq 0.6$, and bulk nature of superconductivity is suppressed by Ni heavy doping. For $x \leq 0.6$, the evolution of the electronic DOS($E_F$) well explains the change in $T_c$, but it cannot explain the disappearance of bulk superconductivity at $x \geq 0.7$. By analyzing the $c/a$ ratio, we revealed a possible transition to collapsed tetragonal phases with a boundary concentration of $x = 0.6–0.8$ by Ni heavy doping. The lattice collapse would affect electronic structure and be negatively linked to superconductivity in $Co_{1-x}Ni_xZr_2$. In addition, the lattice collapse seems to be linked to the appearance of $c$-axis PTE. Since superconductivity in $Co_{1-x}Ni_xZr_2$ would be mediated by phonon, the correlation between axis thermal expansion, emergence of superconductivity, and the lattice collapse transition is one of the notable features of this system. Thus, $Co_{1-x}Ni_xZr_2$ is a suitable platform to study anomalous axis thermal expansion and the method to systematically control the thermal expansion. Furthermore, the relationship between lattice collapse and/or anomalous axis thermal expansion and emergence of superconductivity would provide us with new strategy on exploration of new superconductors.

**Acknowledgements**

The authors thank Md. R. Kasem, A. Yamashita, O. Miura for supports in experiments and D. Louca for fruitful discussion on local structures. The work has been partly supported by JSPS KAKENHI (Grant Number: 21H00151, 21K18834), Tokyo Metropolitan Government Advanced Research (Grant Number: H31-1), TMU strategic fund for young scientists, and TMU strategic fund for multidisciplinary collaboration.

**Author statements**

All the authors declare no competing interests.

**Graphical Abstract**

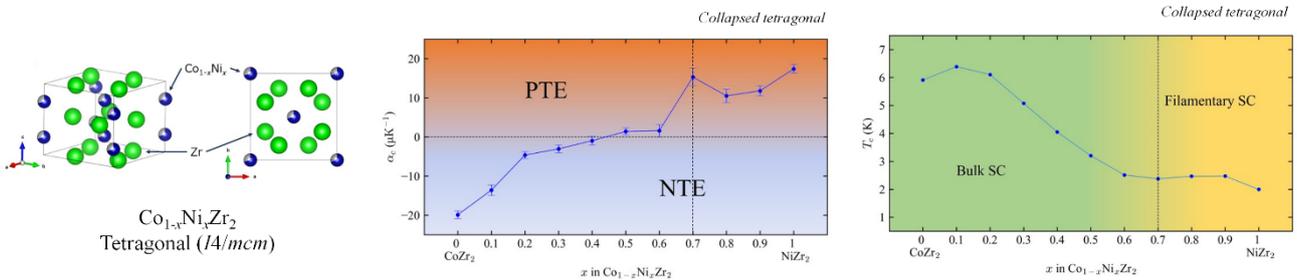

# Supporting materials

### S.1 The XRD patterns for $x$ = 0-1.0.

We measured XRD patterns for all samples. Main peaks could be indexed with a tetragonal $CuAl_2$-type structure ($I4/mcm$), however, some impurity peaks developed as shown in asterisks. These impurity peaks are indicated with an orthorhombic $TrZr_3$ (space group #63) phase [S1]

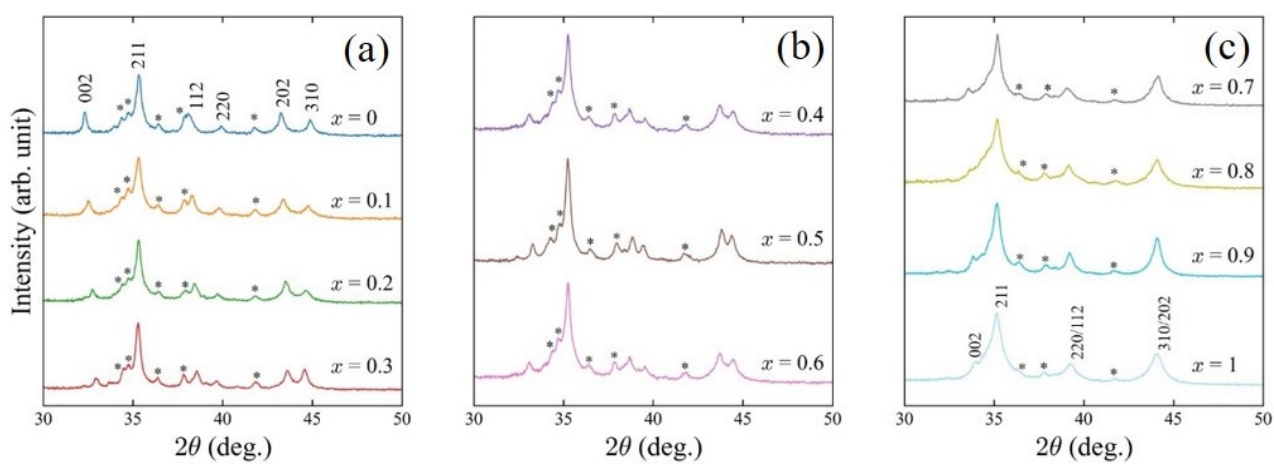

Fig. S1. XRD patterns of $Co_{1-x}Ni_xZr_2$ for $x$ = (a) 0–0.3, (b) 0.4–0.6, (c) 0.7–1.0. Asterisks indicate the impure phases of $TrZr_3$.

### S.2 Thermal expansion results for $x$ = 0.3

We observed the thermal expansion for all samples, here, we show the results of $x$ = 0.3. This sample exhibits $a$-axis, volume PTE and $c$-axis NTE as shown in Supplementary Fig. 2(a)-2(c). Supplementary Figure 2(d) shows High-temperature XRD patterns and the gradual shift of the 002 peak to the higher angle side. This behavior is consistent with the result of $c$-axis contract.

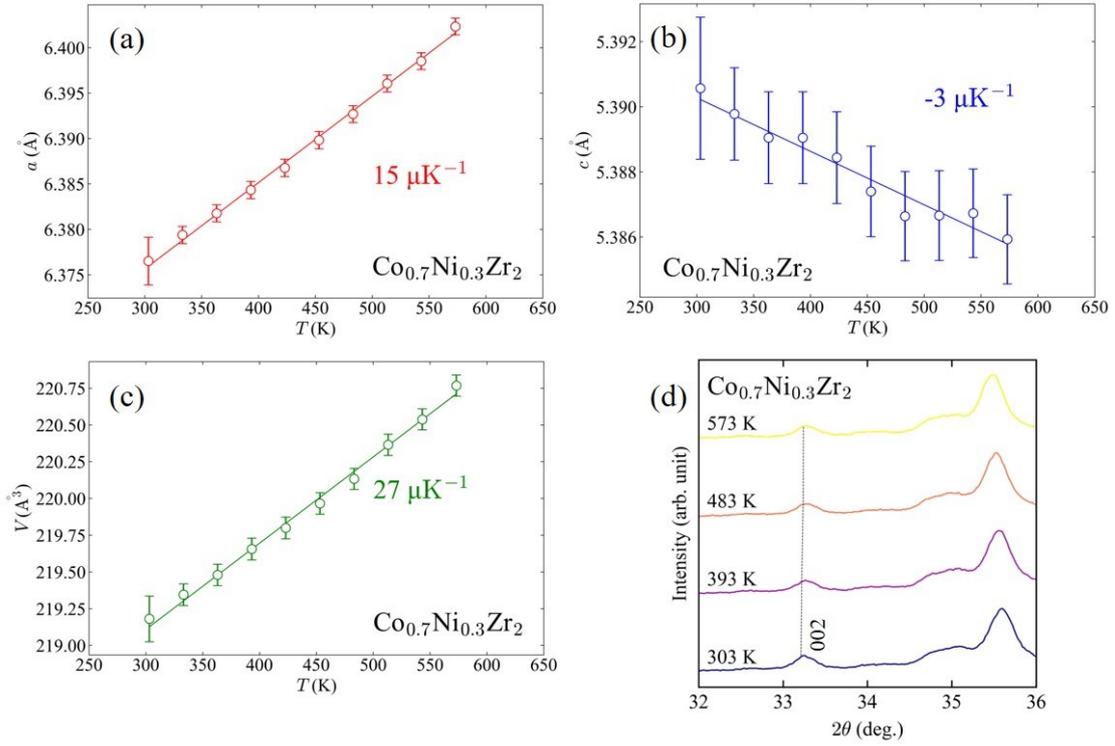

Fig. S2. (a, b, c) The temperature dependence of lattice constants $a$, $c$, and $V$ for $x = 0.3$ sample. The estimated linear thermal expansion constants are $\alpha_a = +14.9(6)$, $\alpha_c = -3(1)$. The volumetric expansion coefficient is $\beta = +27(1)$ μK$^{-1}$ (d) High-temperature XRD patterns on $T = 303, 393, 483, 573$ K for $x = 0.3$ sample. The dashed line in this graph indicates that the 002 peak shifts to the higher angle side.